\documentclass{sig-alternate}
\usepackage{textcomp}
\usepackage[numbers,square]{natbib}
\usepackage{amsfonts}
\usepackage{float}
\makeatletter
\renewcommand\newfloat[3]{\@namedef{ext@#1}{#3}
  \let\float@do=\relax
  \xdef\@tempa{\noexpand\float@exts{\the\float@exts \float@do{#3}}}%
  \@tempa
  \floatplacement{#1}{#2}%
  \@ifundefined{fname@#1}{\floatname{#1}{#1}}{}
  \expandafter\edef\csname ftype@#1\endcsname{\value{float@type}}%
  \addtocounter{float@type}{\value{float@type}}
  \restylefloat{#1}%
  \expandafter\edef\csname fnum@#1\endcsname%
    {\expandafter\noexpand\csname fname@#1\endcsname{}~%
       \expandafter\noexpand\csname the#1\endcsname:}
  \@ifnextchar[{\float@newx{#1}}%
    {\@ifundefined{c@#1}{\newcounter{#1}\@namedef{the#1}{\arabic{#1}}}%
      {}}}
\makeatother
\usepackage{xspace}
\usepackage{algorithmwh}
\usepackage{amssymb}
\usepackage{latexsym}
\usepackage{stmaryrd}
\usepackage{amsmath}
\usepackage{paralist}
\usepackage{array}

\usepackage{tikz}
\usetikzlibrary{decorations}
\usetikzlibrary{automata}
\usetikzlibrary{shapes.multipart,shapes.misc}
\usetikzlibrary{arrows}

\newtheorem{mydef}{Definition}

\newcommand{\uendif}{{\bf end-if}\xspace}

\newcommand*\grant{\textup{\textsf{grant}}\xspace}
\newcommand*\serve{\textup{\textsf{serve}}\xspace}
\newcommand*\request{\textup{\textsf{req}}\xspace}
\newcommand*\clean{\textup{\textsf{clean}}\xspace}
\newcommand{\EmphDefinition}[1]{\textit{\textbf{#1}}}

\newcommand{\bi}{\begin{itemize}}

\newcommand{\ei}{\end{itemize}}
\begin{document}


\title{Long-Term Average Cost in Featured Transition Systems}

%

\author{
\alignauthor
Rafael Olaechea*, Uli Fahrenberg$^{\dagger}$, Joanne M. Atlee*, Axel Legay$^{\dagger}$  \\
      \affaddr{* University of Waterloo, Canada}   \\
      \affaddr{$\dagger$ Inria Rennes, France} \\           
              \email{ Emails: \{reolaech, jmatlee\}@uwaterloo.ca, \{ulrich.fahrenberg, axel.legay\}@irisa.fr} \\
}

\maketitle

\vspace{-5ex}

\begin{abstract}
  A software product line is a family of software products that share
  a common set of mandatory features and whose individual products are
  differentiated by their variable (optional or alternative)
  features.  Family-based analysis of software product lines takes as
  input a single model of a complete product line and analyzes all its
  products at the same time.  As the number of products in a software
  product line may be large, this is generally preferable to analyzing
  each product on its own.  Family-based analysis, however, requires
  that standard algorithms be adapted to accomodate variability.

  In this paper we adapt the standard algorithm for computing limit
  average cost of a weighted transition system to software product
  lines.  Limit average is a useful and popular measure for the
  long-term average behavior of a quality attribute such as
  performance or energy consumption, but has hitherto not been
  available for family-based analysis of software product lines.  Our
  algorithm operates on weighted featured transition systems, at a
  symbolic level, and computes limit average cost for all products in
  a software product line at the same time. We have implemented the
  algorithm and evaluated it on several examples.
\end{abstract}



\section{Introduction}

Many of today's software-intensive systems are developed as a family of related systems (e.g., smart phones, automotive software). In particular, a \emph{software product line (SPL)} is a family of software products that share a common set of mandatory features and whose individual \emph{products} are differentiated by their variable (optional or alternative) \emph{features}. 

Analysis of software product lines can be categorized into family-based or product-based  \cite{Thum:2014:CSA:2620784.2580950}. 
Product-based techniques analyze each possible product (or a sample subset of products) individually, whereas a family-based analysis is performed on a single model that represents all of the products in an SPL.
Thus, family-based approaches avoid some of the redundant computations inherent in product-based analyses; but they require that standard analysis algorithms be adapted to accommodate variability in the SPL model.

We are interested in family-based analyses of quality attributes of software systems (e.g., performance, energy consumption). 
An especially useful analysis of quality attributes, called
\emph{limit average}, computes a long-term average of a quality
attribute for a product.
In this paper, we adapt the limit average algorithm in order to perform a family-based
analysis that computes the limit average for all products in a
software product line.

Our contributions include: 
\begin{itemize}
\item A family-based algorithm that analyzes a model of an SPL and
  computes the limit average for a quality attribute, for all
  products at the same time.
\item An implementation of the family-based algorithm.
\item An evaluation of the speed-up of our family-based approach
  versus the product-based approach.
\end{itemize}

\section{Background}



A \emph{transition system} (TS) is composed of a set of states,
actions, transitions and a set of initial states.  More formally, it is a
tuple $ts = (S, Act, trans, I)$, where $trans\subseteq S \times Act
\times S$ and $I\subseteq S$.
%
%
An \emph{execution} of a transition system is an alternating infinite
sequence of states and actions $\pi=s_0\alpha_1s_1\alpha_2\ldots$ with
$s_0 \in I$ such that  $ (s_i, \alpha_{i+1}, s_{i+1}) \in trans$ for
all $i$. The semantics of a TS (written as $ \llbracket ts \rrbracket $) are given by its set of executions.


A software system may have to satisfy not only \emph{functional
  requirements},  which can be expressed and verified for example through logical properties, but also \emph{quality requirements} such as maximum energy consumption or timing constraints.
Hence transition systems have been extended with weights to model these quality attributes. 
%
%
A \emph{weighted transition system} is thus a tuple
$wts = (S, Act, trans, I, W )$, where $( S, Act, trans, I)$ is a
transition system and $W$ : $ trans \rightarrow \mathbb{R}$ is a
function that assigns real weights to transitions.


\subsection{Limit Average Cost}
\label{sec:LimitAverage}

The \emph{limit average cost} expresses the average of weights in a single infinite execution of a weighted transition system. 
Thus, if the weights represent the consumption of a resource, then the limit average represents the long-term rate of resource consumption along a single (infinite) execution.

Given an infinite execution $\pi=s_0\alpha_1s_1\alpha_2\ldots$ of a
weighted transition system, we define a corresponding infinite
sequence of weights $w(\pi)=v_0v_1\ldots$ where
$v_i=W(s_i\alpha_{i+1}s_{i+1})$.  The limit average of $\pi$ is then
defined to be
\begin{equation*}
  LimAvg(\pi)= \liminf_{n\to\infty}  \frac{1}{n}\sum_{i=0}^{n} v_{i}\,.
\end{equation*}



%
The \emph{maximum (or minimum) limit average} of a weighted transition system is the maximum (or minimum)  limit average over all of its execution traces.
For example, by computing the minimum and maximum limit average of a
weighted transition system whose weights represent energy consumption,
we obtain the best-case and worst-case long-term rates of energy
consumption.

Computation of maximum or minimum limit-average cost is entirely
analogous.  In this paper we focus on maximum limit-average cost, but
everything we do can also be applied to minimum limit-average cost.
The maximum limit average can be computed by a two-phase
algorithm~\cite{Zwick1996343}:
first one computes the set of strongly connected components,
and then for each strongly connected component one identifies the cycle with the highest mean-weight. 
Finally, the mean weight of the maximum mean-weight cycle reachable
from the initial state is the maximum limit average for the weighted
transition system.

A \emph{strongly connected component (SCC)} is a maximal set of nodes in a
graph such that there exists a directed path between every pair of
nodes in the set.
Any cycle in a graph will be contained inside an SCC, hence by
searching for maximum mean-weight cycles in each SCC of a graph we
obtain the maximum mean-weight cycle of the full graph.


The standard algorithm~\cite{%
  Cormen:2001:IA:580470} for computing the SCCs of a graph $G=(V,E)$
performs a depth-first search of the graph and computes for each node
its ``finishing time'' in the depth-first search.
The finishing time $F( v)$ of a node $v$ represents the temporal order
in which the node and all its subsequent neighbors have been fully
explored, and ranges from 1 to $|V|$.
%

The algorithm for computing SCCs then processes the nodes in
decreasing finishing times.
It starts at the node $v$ with $F(v)=|V|$ and computes the set of
nodes that can be reached from $v$ in the transpose of the graph
(i.e.~the graph that has the same nodes and edges but with reversed
edge directions).
These sets of nodes correspond to an SCC.
The algorithm then removes this SCC from the graph and processes the
remaining nodes in decreasing order of finishing times,
%
until each node has been assigned to an SCC.
The SCC algorithm takes time $O(V + E)$.

In order to compute the maximum limit-average cost, we now need to
calculate the highest mean-weight cycle in each SCC.  This is usually
done using Karp's algorithm~\cite{Karp78}.  This algorithm choses an
arbitrary initial state $s_0$ and then iteratively computes a function
$D$ which associates with each state $v$ and each $k\in\{ 0,\dotsc,
n\}$, where $n$ is the size of the SCC, the maximal weight of a path
of length $k$ from $v$ to $s_0$.  By Karp's theorem~\cite{Karp78}, the
weight of the maximal mean-weight cycle is then given as $\max_v
\min_{ k< n} \frac{ D[ n, v]- D[ k, v]}{ n- k}$.

\subsection{Weighted  Featured Transition Systems}


A \emph{feature model}~\cite{Kang90Short} is used to configure a
software product line.
It represents the set of valid  products. 
For our purposes a feature model is used exclusively to distinguish
between valid and invalid products, hence it is
%
a tuple $d=(N, px)$, where $N$ is the set of features and
$px\subseteq \mathcal P(N)$ is the set of products.
%
%
Here $\mathcal P( N)$ denotes the power set of $N$; an individual
software product is thus composed of a set of features.
%



A transition system represents the behavior of a single software product.
In order to analyze all the products of a software product line at the
same time, Classen \emph{et
  al.}~\cite{Classen:2013:FTS:2516408.2516486} have introduced
featured transition systems which compactly represent the behavior of
all the products of a software product line.

A boolean feature variable represents the presence or absence of a feature in a software product.
A product is then represented by an assignment of values to all feature variables (true if the feature is present in the product, false if not). 
Hence we can represent a set of products by a \emph{boolean feature
  expression} - that is,  a  boolean formula over feature variables,
whose solutions represent the set of products.  We denote by
$\mathbb{B}(N)$ the set of such feature expressions.

A \emph{featured transition system} annotates each transition with a
boolean feature expression, which corresponds to the set of products
whose transition system include that transition.
It is thus a tuple $fts = (S, Act, trans, I, d, \gamma)$, where
$( S, Act, trans, I)$ is a transition system, $d = (N, px)$ is a
feature model, and $\gamma$ : $trans \rightarrow \mathbb{B}(N)$
labels each transition with a feature expression.

Therefore FTSs unify the transition systems of all products in a
product line into a single annotated transition system.
The featured transition system provides a 150\% model of all
products' states and transitions -- that is, it
includes more transitions and states than required for each individual
product.

The transition system for each specific software product can be
derived by removing all annotated transitions whose feature expression
is not satisfied by the product's feature-variable assignment.
%
%
%
This transition system contains all the states of the FTS and all the transitions whose feature expressions evaluate to true under the software product. 
Formally, the \emph{projection} of an FTS $fts$ to a product
$p\in \llbracket d \rrbracket$, noted $fts_{|p}$, is the TS
$ts = (S, Act, trans', I)$ where
$trans' = \{ t \in trans\; | \; p \vDash \gamma(t') \}$.



A featured transition system can be extended with weights on transitions in the same way that transition systems can,
in which case each product of the software product line is represented by a weighted transition system.
Then we can compute the maximum limit average for each product of the
software product line.  A \emph{weighted featured transition system}
(WFTS) is thus a tuple $wfts = (S, Act, trans, I, d, \gamma, W)$,
where $( S, Act, trans, I, d, \gamma)$ is an FTS and
$W : trans \rightarrow \mathbb{R} $ is a function that annotates
transitions with weights.


A WFTS can be projected for a specific product into a weighted
transition system,
analogously to FTS projection as above: the \emph{projection} of a WFTS $wfts$ to a product
$p\in \llbracket d \rrbracket$, denoted $wfts_{|p}$, is the WTS
$wts = (S, Act, trans', I, W)$ where
$trans' = \{ t \in trans\; | \; p \vDash \gamma(t) \}$.


\section{Motivating Example}
\label{se:taxi}

\begin{figure}[tbp]
  \centering
  \begin{tikzpicture}[->, >=stealth', font=\footnotesize, x=1.5cm,
    y=1cm, state/.style={shape=rounded rectangle, draw, initial
      text=, initial distance=2ex, inner sep=.5mm, minimum size=3mm},
    every node/.style={sloped, above}]
    \node[state] (Pe) at (0,0) {Pickup-ext};
    \node[state] (P1) at (0,-2) {Pickup-1};
    \node[state] (P2) at (0,-4) {Pickup-2};
    \node[state] (PA) at (0,-6) {Airport-P};
    \node[state] (Re) at (4,0) {Release-ext};
    \node[state] (R1) at (4,-2) {Release-1};
    \node[state] (R2) at (4,-4) {Release-2};
    \node[state] (RA) at (4,-6) {Airport-R};
    \path (P1) edge node[pos=.75, below] {$40(3)$} (RA);
    \path (PA) edge node[pos=.25, below] {$50(3)$} (R1);
    \path (P2) edge node[pos=.73, below] {$35(2)$} (RA);
    \path (PA) edge node[pos=.27, below] {$45(2)$} (R2);
    \path (Pe) edge[dotted] node[pos=.89] {$50(4)$} (RA);
    \path (PA) edge[dotted] node[pos=.11] {$60(4)$} (Re);
    \path (R1) edge[red] node {$-2$} (P1);
    \path (R2) edge[red] node {$-2$} (P2);
    \path (RA) edge[red] node {$-5$} (PA);
    \path (Re) edge[red!70!black, dotted] node[red] {$-2$} (Pe);
    \path (P1) edge node {$15$} node[below, blue] {$S$} (P2);
    \path (R2) edge node[below] {$15$} node[above, blue] {$S$} (R1);
    \path (Pe) edge[dotted] node {$15$} node[below, blue] {$S$} (P1);
    \path (R1) edge[dotted] node[below] {$15$} node[above, blue] {$S$} (Re);
    \path (P1) edge node[pos=.35] {$30$} node[blue, pos=.35, below] {$T$} (R2);
    \path (P2) edge node[pos=.65] {$30$} node[blue, pos=.65, below] {$T$} (R1);
    \path (Pe) edge[dotted] node[pos=.3] {$30$} node[blue,
    pos=.3, below] {$T$} (R1);
    \path (Pe) edge[dotted] node[pos=.38] {$30$} node[blue,
    pos=.38, below] {$T$} (R2);
    \path (P1) edge[dotted] node[pos=.7] {$30$} node[blue,
    pos=.7, below] {$T$} (Re);
    \path (P2) edge[dotted] node[pos=.62] {$30$} node[blue,
    pos=.62, below] {$T$} (Re);
  \end{tikzpicture}
  \caption{
    \label{fi:taxishuttle}
    Taxi-shuttle example.  In addition to the feature guards shown,
    all \emph{dotted} transitions are guarded by the feature $L$.  The
    notation ``X(Y)" on transitions to and from the airport
    indicates transitions with weight X and length Y. If ``(Y)" is omitted that means transition has length of one. States refer to the location the vehicles is currently located at.}
\end{figure}
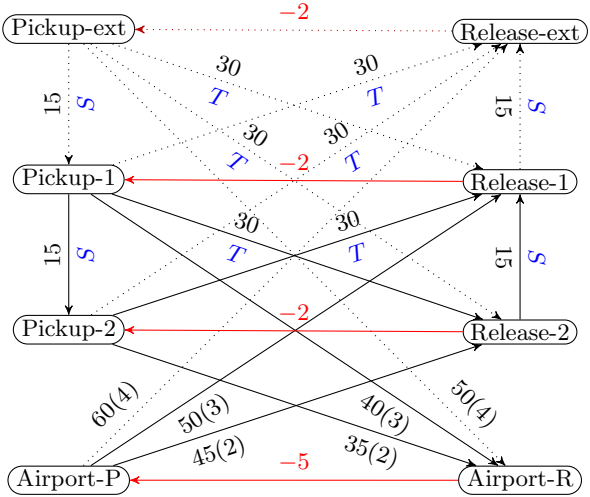

Figure~\ref{fi:taxishuttle} shows an (artificial) example of a
combined taxi and shuttle service.  There are three pickup and release
locations in the city, one of which is available only when the car has
an extra license (feature $L$).  Additionally, passengers can be
picked up and released at the airport.  Taxi service (feature $T$) is
available within city locations, not just for transportation to and
from the airport.  The shuttle service (feature $S$) allows a vehicle to pick up
passengers at several pickup locations before delivering them to the
airport, or to pickup passengers for several different city locations
at the airport.

The weights on the transitions show their cost; positive numbers are
income for the driver, negative numbers are expenses.  To model the
fact that trips to the airport take longer time than trips in the
city, the transitions to and from the airport have length $2$ (from
the second pickup point), $3$ or $4$.  In practice we will
model this by inserting extra states and transitions of weight $0$.

The example has thus three features, $S$, $T$ and $L$, giving rise to
eight products: $\emptyset$, $\{ L\}$, $\{ S\}$, $\{ T\}$,
$\{ L, S\}$, $\{ L, T\}$, $\{ S, T\}$, and $\{ L, S, T\}$.  An
interesting problem is to compute the maximal income for the driver,
depending on the product; the maximum limit average cost is a
reasonable approximation of this maximal income.

A product-based analysis reveals that regardless of the feature
selection, the transition system always has precisely one SCC. In product $p= \emptyset$ there are two cycles:
\begin{gather}
  \text{Airport-P $\to$ Release-1 $\to$ Pickup-1 $\to$} \qquad\qquad
  \notag\\
  \qquad\qquad \text{$\to$ Airport-R $\to$
    Airport-P} \label{ex:taxi-A1A}
  \\
  \text{Airport-P $\to$ Release-2 $\to$ Pickup-2 $\to$} \qquad\qquad
  \notag\\
  \qquad\qquad \text{$\to$ Airport-R $\to$
    Airport-P} \label{ex:taxi-A2A}
\end{gather}
Their mean weights are $10.38$ and $12.17$, respectively (round\-ed to
two places), hence cycle~\eqref{ex:taxi-A2A} between the airport and
city location 2 provides the maximal income.

In $p=\{ L\}$ there are three cycles: the cycles listed above plus a third:
\begin{gather}
  \text{Airport-P $\to$ Release-ext $\to$ Pickup-ext $\to$}
  \qquad\qquad \notag\\
  \qquad\qquad \text{$\to$ Airport-R
    $\to$ Airport-P} \label{ex:taxi-AeA}
\end{gather}
But the mean weight of the third cycle is only $10.30$, so cycle~\eqref{ex:taxi-A2A} is
still the most profitable.

In $p=\{ S\}$, there are five cycles: in addition to cycles~\eqref{ex:taxi-A1A}
and~\eqref{ex:taxi-A2A} above, there are three other cycles:
\begin{gather}
  \text{Airport-P $\to$ Release-2 $\to$ Release-1 $\to$} \qquad\qquad
  \notag\\
  \qquad\qquad \text{$\to$ Pickup-1 $\to$ Airport-R $\to$ Airport-P}
  \\
  \text{Airport-P $\to$ Release-1 $\to$ Pickup-1 $\to$} \qquad \qquad
  \notag\\
  \qquad\qquad \text{$\to$ Pickup-2 $\to$ Airport-R $\to$ Airport-P}
  \\
  \text{Airport-P $\to$ Release-2 $\to$ Release-1 $\to$ Pickup-1
    $\to$} \quad \qquad
  \notag\\
  \qquad\qquad \text{$\to$ Pickup-2 $\to$ Airport-R $\to$
    Airport-P} \label{ex:taxi-A212A}
\end{gather}
Their mean weights are $11.63$, $11.63$, and $12.88$, respectively;
hence for a purely shuttle product, cycle~\eqref{ex:taxi-A212A}, which picks up
and releases passengers at both city locations, is most profitable.

Similar analyses can be done for the other five products. However, a
family-based analysis that computes SCCs and maximum mean-weight
cycles for all products at once would be preferable.  We will come
back to this example in Section~\ref{se:implex}.

\section{Family-Based Limit Average \\ Computation}

We want to compute the maximum limit average cost for each product in a software product line. 
We propose a family-based algorithm that re-uses partial computation results that apply to multiple products. 
The algorithm starts by computings SCCs (subsections
\ref{sec:finishing}
and \ref{sec:scc}) and then for each SCC it computes the SCC\textquotesingle{}s maximum mean
cycle
(subsection \ref{sec:cyclescc}).
 
\begin{figure}[tbp]
  \centering
  \begin{tikzpicture}[->, >=stealth', font=\footnotesize, x=2.5cm,
    y=1cm, state/.style={shape=circle, draw, initial text=, initial
      distance=2ex, inner sep=.5mm,minimum size=3mm}]
    \node[state, initial] (0) at (0,0) {$0$}; 	
    \node[state] (1) at (2,0) {$1$};		%
    \node[state] (2) at (0, -2) {$2$};
    \node[state] (3) at (2,-2) {$3$};

    \path (0) edge node[below] {\request~/~0} (1);
    \path (1) edge node[right] {\grant~/~0 } (3);    
    \path (3) edge node[right, pos=.5] {\;\serve~/~0} (0);
       
      \path (2) edge [loop below, green!50!black] node {\grant $[G]$~/~-1} (2);

          \path (2) edge[out=150, in=210, blue] node[left] { \clean
      $[A]$~/~0} (0);
      
    \path (0) edge[purple] node[right, pos=.7] {\grant $[G
      \lor A]$~/~-1} (2);
    \path (2) edge[purple] node[below] {\request $[G \lor A]$~/~0} (3);    
     \end{tikzpicture}
  \caption{
    \label{fi:ex3}
    WFTS which implements several grant/ request scenarios}
\end{figure}
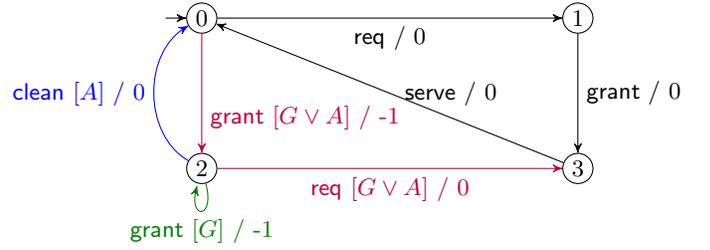
 
In order to illustrate the family-based SCC computation, we introduce
another example.
Consider three solutions to the problem of an arbiter granting access
to a shared resource, modeled as a WFTS in Fig.~\ref{fi:ex3}.
One solution involves granting access only after a request has been
received: this will be the solution implemented by the basic system
without the optional features $A$ or $G$.
An alternative solution is to always grant access, whether a request
exists or not. This is implemented by the product with 
feature $G$.
A third option is to alternate between granting access and not
granting access, implemented by the product with 
feature $A$.

Each of these solutions satisfies the functional requirements of the
system,
namely that a request is always granted.
However the user may prefer one solution over another: for example she
might want to minimize the number of unnecessary grants.
These preferences are encoded as weights on the transitions, such that
every time a grant is given when not needed, or when a request has to
wait before being served, a penalty of $-1$ is given.
%
%

\subsection{Symbolic Finishing Times}
\label{sec:finishing}

The algorithm for computing SCCs of a graph depends on the finishing times of states in a depth-first search.  
However a featured transition system represents a set of transition
systems, each with a different set of transitions, which can give rise
to a different set of depth-first finishing times for its states.
For example the basic product in Fig.~\ref{fi:ex3} (without feature
$A$ nor $G$) would have the following finishing times of states:
\begin{equation*}
  F(s_3)=1, F(s_1)=2, F(s_0)=3, F(s_2)=4\,,
\end{equation*}
whereas in any product that includes feature $A$, state $s_0$ has the
highest finishing time:
\begin{equation*}
  F(s_3)=1, F(s_1)=2, F(s_2)=3, F(s_0)=4\,.
\end{equation*}
Hence to adapt the SCC algorithm to featured transition systems, we
construct a \emph{tree} that symbolically represents all the
possible finishing times of states.

Each path in such a \emph{symbolic finishing-times tree} from the root
to a leaf node represents a unique set of finishing times for the
states in a featured transition system.
The tree is annotated with feature-expression labels on edges,
associating products with states' finishing times.
Specifically, a tree node representing state $s$ at level $d$ in the
tree means that the finishing time of state $s$ is $|S|-d+1$ in \emph{all
products} that satisfy the
feature expressions along the path from the root
to the node.

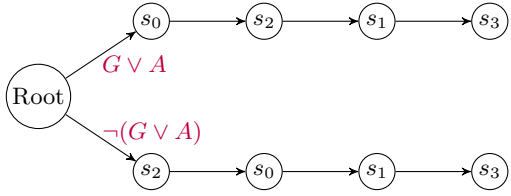
\begin{figure}[tbp]
  \centering
  \begin{tikzpicture}[->, >=stealth', font=\footnotesize, x=1.5cm,
    y=1cm, state/.style={shape=circle, draw, initial text=, initial
      distance=2ex, inner sep=.5mm,minimum size=3mm}]
    \node[state] (0) at (-2,0) {Root}; 	
    \node[state] (1) at (-1,1.0) {$s_0$};		%
     \node[state] (2) at (-1,-1.0) {$s_2$};		%
    \node[state] (3) at (0.0,1.0) {$s_2$};		%
     \node[state] (4) at (0.0,-1.0) {$s_0$};		%
    \node[state] (5) at (1.0,1.0) {$s_1$};		%
     \node[state] (6) at (1.0,-1.0) {$s_1$};		%
    \node[state] (7) at (2.0,1.0) {$s_3$};		%
     \node[state] (8) at (2.0,-1.0) {$s_3$};		%
     
    \path (0) edge node[right, purple, pos=.3] {\;$G \lor A$} (1);
     \path (0) edge node[right, purple, pos=.4] {$\neg(G \lor A)$} (2);

     \path (1) edge node[] {} (3);
     \path (3) edge node[] {} (5);
     \path (5) edge node[] {} (7);

     \path (2) edge node[] {} (4);
     \path (4) edge node[] {} (6);
     \path (6) edge node[] {} (8);
                   
     \end{tikzpicture}
  \caption{
    \label{fi:ExampleSymbolicFinishingTimes}
   Symbolic finishing-times tree for the FTS from Fig.~\ref{fi:ex3}}
\end{figure}

For example, the WFTS from Fig.~\ref{fi:ex3} gives rise to the
symbolic finishing-times tree shown in
Fig.~\ref{fi:ExampleSymbolicFinishingTimes}.
This tree assigns one set of finishing times for all products that
contain either feature $G$ or $A$, and another set of finishing times
for products that contain neither feature.
 
\begin{mydef}\label{def:SymbolicPostCondition}
  Let fts be a featured transition system. A \EmphDefinition{symbolic
    finishing-times tree} for fts is composed of a tree $T=(V, E)$ of
  height $n=|S|$, a node labelling function
  $\ell_v: (V \setminus root) \rightarrow S$ and a function
  $\ell_e: E \rightarrow \mathbb{B}(N)$ which labels each edge with a
  feature expression.  The tree $T$ satisfies the following
  conditions:
\begin{compactitem}
  \item All leaf nodes are at level $|S|$ of the tree.
  \item For any path $v_0, \ldots, v_n$ from the root to a leaf node,
    each node $v_i$ is mapped to a unique state:
    $\forall i, j \in \{1 \ldots n\}, i \neq j: \ell_v(v_i) \neq
    \ell_v(v_j) $.
    A path from the root to a leaf node represents a set of products
    that share the same finishing times for its nodes.
  \item The feature expressions of outgoing edges from a node are
    disjoint: $\forall (u, v),$
    $(u, w) \in E, w \neq v : \llbracket \ell_e((u,v)) \rrbracket \cap
    \llbracket \ell_e((u,w)) \rrbracket = \emptyset$.
  \item For any product $p$ and level $i$, there exists a (necessarily
    unique) path $v_0, \ldots, v_i$ from the root to a node in level
    $i$ such that the product $p$ is contained in the conjunction of
    the feature expressions along the edges of the path:
    $\forall p \in \llbracket d \rrbracket, i \in \{1, \ldots, n\} :
    \exists$~a~path
    $v_0, \ldots, v_i \; : p \in \bigcap_{j=0}^{i-1} \llbracket
    \ell_e((v_j, v_{j+1})) \rrbracket$.
  \item For any product $p$, level $i$, and the unique path from the
    root $v_0, \ldots, v_i$ such that
    $p \in \bigcap_{j=0}^{i-1} \llbracket \ell_e((v_j, v_{j+1}))
    \rrbracket $,
    the finishing times in the projection $fts_{|p}$ of the states
    $\ell_v(v_1), \ldots,$ $\ell_v(v_i)$ are $n, \ldots, n- i + 1$,
    respectively.
  \end{compactitem}
\end{mydef}
 
 

The symbolic finishing-times tree is built in two phases.
In the first phase (performed by Alg.~\ref{alg:dfs-fts}), a symbolic
depth-first search explores all states of an FTS and computes a
temporal ordering for when a state and all of its neighbors are
explored, depending on feature expressions.
The second phase (shown in Alg.~\ref{alg:bfs-tree}) uses this
information to construct a symbolic finishing-times trees in a
breadth-first manner.


In Alg.~\ref{alg:dfs-fts},
%
unlike in a standard depth-first algorithm, states are not marked as
visited
by a boolean flag, but instead with a feature expression representing
\emph{under which set of products} they have been visited.
Hence Alg.~\ref{alg:dfs-fts} stores and updates an array \textbf{White} of boolean formulas: representing the products for which a state has not been explored 

\begin{algorithm}[tb]
\caption{
  \label{alg:dfs-fts}
  {\bf Featured transition system depth first search}}
\begin{code}
\uln \> {\bf Procedure } DFS-Fts (G)  \\
\uln \> \ubegin \\
\uln \> \>  \uforeach  u $\in V[G]$  \\
\uln \> \>  \> color[u][White] $ \uassign \top $   \\
\uln \> \>  time $ \uassign $ 0  \\
\uln \> \>  \uforeach  u $\in V[G]$  \\
\uln \> \>  \> \uif  color[u][White] is satisfiable   \\
\uln \> \>  \>  \> DFS-Fts-Visit(u, color[u][White]) \\
\uln \> \>  \> \uendif  \\
\uln \> \uend \\
\uln \> {\bf Procedure } DFS-Fts-Visit(u, $\lambda$)  \\
\uln \> \ubegin \\
\uln \> \> Exploring $ \uassign$ color[u][White] $\wedge \; \lambda $  \\
\uln \> \> color[u][White] $ \uassign$  color[u][White]  $ \wedge \neg \lambda $  \\
\uln \> \>  \uforeach  (u, v, $\lambda'$)  $\in E[G]$  \\
\uln \> \>  \> NextFExp $\uassign \lambda' \wedge \lambda $  \\
\uln \> \>  \>  \uif  (color[v][White]  $\wedge$ NextFexp) is sat.   \\
\uln \> \>  \>  \>  DFS-Fts-Visit(v, NextFexp)  \\
\uln \> \>  \> \uendif  \\
\uln \> \> time $ \uassign $ time + 1 \\
\uln \> \> O[u][Exploring] $ \uassign $ time  \\
\uln \> \uend
\end{code}
\end{algorithm}

Algorithm~\ref{alg:dfs-fts} starts by initializing array White to true (all products)  for each state (lines 3-4). 
It then iterates over all states, and for each state that has not been
fully explored, it calls the subroutine DFS-Fts-Visit with that state and the feature expression representing the  set of unexplored products as parameters (lines 6-8). 

The subroutine DFS-Fts-Visit starts by updating (reducing) the set of
unexplored products for its given state (line 13).
Then it iterates over each outgoing edge and checks if there are
products for which the target state has not been explored, i.e.~if
color[v][White] $\land \lambda' \land \lambda$ is satisfiable (line 17).
If so, then it recursively calls itself to explore the destination
state.
%
Finally, once all outgoing edges
have been explored, it sets the finishing time for the given state and
feature expression to the current time counter and increments this
counter.
  
\begin{algorithm}[tbp]
\caption{
  \label{alg:bfs-tree}
  {\bf Building a symbolic finishing-times tree for  an  FTS}}
\begin{code}
\uln \> {\bf Procedure } ComputeTreeBfs(O, OInv)  \\
\uln \> \ubegin \\
\uln \> \> Q $\uassign $ Empty Queue  \\
\uln \> \> T $\uassign$ New  Tree() \\
\uln \> \>  T.root.maxO $\uassign$  $|$domain(O)$|$  \\
\uln \> \> Q.add(T.root) \\
\uln \> \>  \uwhile ($\neg $Q.isEmpty())  \\
\uln \> \> \>  Node $\uassign$ Q.pop() \\
\uln \> \>  \>   $\lambda_{1} \uassign $ \makebox[0pt][l]{FeatureExpressionFromRoot(Node)} \\
\uln \> \>  \>  max $\uassign $  Node.maxO \\
\uln \> \>  \>    notChildren $\uassign $  $ \top $  \\ 
\uln \> \>  \>     j $\uassign $  max -1 \\
\uln \> \>   \>   $\uwhile\; (j > 0)$ \\
\uln \> \>     \> \>  u, $\lambda$  $\uassign$  OInv(j) \\
\uln \> \>     \> \> \uif ($\lambda\; \wedge \; $notChildren$\; \wedge \; \lambda_{1} $   is sat.) \\
\uln \> \>     \> \> \> NewNode $\uassign$  CreateNode(Node, \\
\hspace*{16em} u, $ \lambda \; \wedge \;$notChildren) \\
\uln \> \>     \> \> \> Q.add(NewNode) \\
\uln \> \>     \> \> \> notChildren $\uassign$ \makebox[0pt][l]{notChildren$\;\wedge \; \neg  \lambda$} \\ 
\uln \> \>    \> \>  \uendif \\
\uln \> \>    \> \>  j $\uassign  j-1$  \\
\uln \>  \> \ureturn T \\ 
\uln \> \uend \\ 
\uln \> {\bf Procedure } CreateNode(ParentNode, State, $\lambda$)  \\
\uln \> \ubegin \\
\uln \> \> NewNode $\uassign$ New  Node() \\
\uln \> \>  ParentNode.add(NewNode) \\
\uln \> \>  StateLabel(NewNode) $\uassign$ State \\
\uln \> \>  EdgeLabel(ParentNode, NewNode) $\uassign$ $\lambda$ \\
\uln \> \uend \\
\uln \> {\bf Procedure } FeatureExpressionFromRoot(Node)  \\
\uln \> \ubegin \\
\uln \> \> \uif Node $=$ T.root \\
\uln \> \> \>  \ureturn $\top$ \\
\uln \> \> \uelse \\
\uln \> \> \> \ureturn \makebox[0pt][l]{EdgeLabel(Parent(Node), Node)
  $\wedge$} \\  
\hspace*{6.5em} FeatureExpressionFromRoot(Parent(Node)) \\
\uln \> \> \uendif \\
\uln \> \uend
\end{code}
\end{algorithm}

Once the feature-based depth-first ordering of states has been
computed, this data can be used to construct the symbolic
finishing-times tree for the FTS.  We do this by iterating over the
states in reverse finishing order, recursively adding a new child to a
tree node whenever a new pair $( s, \lambda)$ is found for which
$\lambda$ is not contained in the disjunction of the feature
expressions along the edges to the other children.  The procedure is
shown in Alg.~\ref{alg:bfs-tree}.

The algorithm starts by initializing a tree T with an empty root node and adding it to a queue of tree nodes to explore  (lines 3-6). 
It then enters a loop  where it processes  tree nodes from the queue and computes all their children (lines 7-20).
%
%

In order to identify all children of a tree node, the algorithm iterates over order numbers lower than than the maximum order number stored in the tree node in decreasing order (lines 13-20).
It searches  for pairs of states and feature expressions $(s, \lambda) =$~OInv($i$) such that the feature expression ($\lambda$) combined 
with the negation of all other edges leaving the tree node is satisfiable (line 15-19). 
If the feature expression is satisfiable, then it adds the new
children  to the tree (line 16) and updates the expression
representing the negation of all edges leaving the tree node (line 18). 
It records the order number in the tree node and  then adds the new tree node to the queue (line 17).
After all children for a tree node have been identified and added, any tree nodes remaining in the queue are processed (line 7).

\subsection{Strongly Connected Components of a \\ Featured Transition System}
\label{sec:scc}

After building the symbolic finishing-times tree,
we use this tree to compute the SCCs of an FTS.
We adapt the standard algorithm for computing SCCs (see
Sect.~\ref{sec:LimitAverage}) by replacing the single set of
finishing times by the symbolic finishing-times tree.
Hence we no longer compute a single set of SCCs, but instead compute
one such set for each path from the root to a leaf node in the
tree.
This adaptation is necessary as the ``finishing times'' of states in
an FTS depend on which features are present in a given product.

%

We explore each path from the root to a leaf node of the symbolic finishing-times tree. 
In the standard SCC algorithm, a boolean array keeps track of which
states have been assigned to an SCC.
In our case, we use an array of feature expressions representing
\emph{for which products} a state has been assigned.
%
%
The algorithm to compute the symbolic SCCs is shown as
Alg.~\ref{alg:scc}.  It uses a subroutine VisitDFS-For-SCC which we
show as Alg.~\ref{alg:main-helper}.

%
%
%

\begin{algorithm}[tbp]
\caption{\small {\bf Computing strongly connected components for an FTS given a symbolic finishing-times tree}. }
\begin{code}
\uln \> {\bf Procedure } SymbolicSCC  \\
\uln \> {\bf Input:} T, NodeLabel, EdgeLabel: a symbolic
\\ \hspace*{16em} finishing-times tree
\\ 
\uln \> {\bf Output:} $RC$: A function from tree nodes 
\\ \hspace*{16em} to symbolic SCCs  \\ 
\uln \>\ubegin\\
\uln \>  \> NodesToExplore $\uassign $ empty stack of triplets of \\
\hspace*{7.5em} tree nodes, states and feature expressions \\
\uln \> \>  ReachabilityStack $\uassign $ empty stack of  \\
\hspace*{16em} mappings $S\to \mathbb{B}( N)$ \\
\uln \>  \>  {\bf For each}\xspace   $e=($Root(T), u) $ \in E(T)$ \\
\uln \>  \> \> $ R'  \uassign \{\} $ \\
\uln \>  \> \> $ \lambda_0  \uassign $  EdgeLabel(e)  \\ 
\uln \> \> \> $ s_0 \uassign $ NodeLabel(u) \\ 

\uln \> \>  \> NodesToExplore.push((u,$\; s_0,   \lambda_{0}$)) \\
\uln \> \>  \> ReachabilityStack.push($R'$) \\

\uln  \> \>\>  \uwhile NodesToExplore $\neq []$ \udo  \\

\uln \> \> \> \> u, s, $  \lambda  \uassign$ NodesToExplore.peek()\\

\uln \> \> \> \> Visited(u)$\uassign$ True \\


\uln \> \> \> \>  \uif $\lambda\land \neg R'( s)$ is satisfiable \\

\uln \> \> \>  \>\>	$ RC$(u) $ \uassign$ VisitDFS-For-SCC \\ 
\hspace*{17.5em} $(s, \lambda \wedge \neg R'(s), R')$ \\

\uln \> \> \>  \>\>	$R'  \uassign R' \cup  RC$(u)  \\

\uln \>\> \> \> \uendif \\

\uln \>\> \> \>  Take v in Children(u) with   \\
\hspace*{17.5em} Visited(v)=False  \\

\uln \>\> \> \> \uif no such v exists: \\

\uln \>\> \> \> \> NodesToExplore.Pop();\\
\uln \>\> \> \> \> $R'  \uassign$ ReachabilityStack.Pop() \\

\uln \>\> \> \> \uelse \\
\uln \>\> \> \> \>  $\lambda' \uassign$ EdgeLabel(u,v) \\

\uln \>\> \> \> \>  NodesToExplore.push((v,  \\
		\> \> \>  \>\> \>\>  NodeLabel(v), $ \lambda \wedge \lambda' $))\\

\uln \>\> \> \> \>  ReachabilityStack.push($R'$) \\

\uln \>\> \> \> \uendif \\

\uln \>\>   \ureturn $RC$ \\
\uln \> \uend
\end{code}
\label{alg:scc}
\end{algorithm}

The output of Alg.~\ref{alg:scc} is a \emph{symbolic SCC tree}.  Its
tree structure is the same as the symbolic finishing-times tree, but
now the tree nodes are labeled with mappings from $S$ to
$\mathbb{B}( N)$, representing for which products a given state is
assigned to a particular SCC.

Algorithm~\ref{alg:scc} starts by successively exploring each outgoing
edge from the root of the tree (line 7).  It then adds a triplet
consisting of the child of the root node, along with its state and
feature expression labels, to a stack of nodes to explore (lines
9-11).


The algorithm then enters a loop where elements of the stack are
processed (lines 13-28), which corresponds to a depth first
exploration of the finishing times tree.
A triplet of tree node, state and feature expression is peeked
from the stack (without being popped).
The feature expression is compared to $R'(s)$ which contains the set
of products for which the given state is already assigned to an SCC,
and if it is not contained in $R'(s)$, then a new symbolic SCC is
computed by calling VisitDFS-For-SCC (line 16-17).
The set of products assigned to an SCC for each state is then updated.

After processing the current tree node, the algorithm looks for a
child that has not been explored (line 20).
If no such child exists, then the current element is popped from the
stack, otherwise a triplet is built from the child node, its state
label and the feature expression labelling the edge to it and pushed
to the stack of nodes to explore (lines 25-27).
The algorithm continues processing triplets in the stack until it is
empty and the complete finishing-times tree has been explored.
%
 
%


%

The procedure VisitDFS-For-SCC computes the set of states which are
reachable from a given state $s$ in the transpose of the input DFS,
parameterized by feature expressions.  This is inspired by the
symbolic reachability algorithm
of~\cite{Classen:2013:FTS:2516408.2516486}, except that here we
exclude states from the search which have already been assigned to
previous SCCs.  The procedure
is shown as Algorithm~\ref{alg:main-helper}. 


\begin{algorithm}[tp]
\caption{\small {\bf Reachability computation for the transpose of an FTS,  excluding  states already assigned to an SCC.} }
\begin{code}
\uln \> {\bf Procedure }  VisitDFS-For-SCC($s_0$, $\lambda_0$, $R'$)   \\
\uln \> {\bf Inputs:} \>\> $s_o$: initial state of the SCC \\
\> \> \> \> $\lambda_0$: initial feature expression of the SCC \\
\> \> \> \> $R': S \rightarrow \mathbb{B}(N)$: the (symbolic) set of \\
\hspace*{10em} states which are  already  assigned \\
\hspace*{10em} to an SCC  \\
\uln \> {\bf Output:} $ R : S \rightarrow \mathbb{B}(N) $ \\ 
\uln \>\ubegin\\


\uln \>  \> $ R \uassign \{ (s_0, \lambda_{0}) \}$ \\

\uln \> \>  Stack.push$((s_0,   \lambda_{0}))$ \\

\uln  \> \> \uwhile Stack $\neq []$ \udo  \\

\uln \> \> \> 	(s, px) $\uassign$ Stack.peek()\\

\uln \> \> \> new $ \uassign \{ (s', px')  \in$ Post$(s, px) \mid $ \\
\hspace*{14em} $px'  \not\subseteq   R(s')\cup R'(s')\}$ \\

\uln \> \>\> \uif   new =  $\emptyset $ \uthen \\

\uln \> \>\> \> 	pop(Stack); \\

\uln \> \>\> \uelse \\ 

\uln \> \>\> \>  Take  $ (s',  px') \in$ new  \\ 

\uln \> \>\> \>    $ R(s') \uassign R(s') \cup   (px' \cap \neg R'(s))   $  \\ 

\uln \> \> \> \>  Stack.push$((s',  px' \cap \neg R'(s) ))$ \\

\uln \> \> \> \uendif \\

\uln \> \>  \ureturn $R$ \\  

\uln \>\uend
\end{code}
\label{alg:main-helper}
\end{algorithm}

Algorithm~\ref{alg:main-helper}
takes as input an initial state, feature expression and symbolic set
of excluded states, and computes the symbolic set of states reachable
from the initial state and feature expression without going through
any of the excluded states.
%
This modified reachability algorithm returns a symbolic set of states: a
mapping of states to feature expressions representing the set of
states reachable under a given product.

Algorithm~\ref{alg:main-helper} starts by initializing an empty
reachability relationship $R$ with the initial state and feature
expression and pushing the initial state and feature expression into a
stack (line 7-9). It then enters a loop where it processes elements of
the stack until the stack is empty (lines 10-20).

The algorithm peeks at the top element of the stack and computes the set of its successors that are not a member of either  $R$ or of excluded states $R'$ (lines 11-12).  
If this set of new elements is empty then it pops the top element of stack (lines 14-15). 
Otherwise it takes a state and feature expression that is a new element, updates $R$ with it and pushes the new element into the stack (lines  17-19).
It then continues  processing elements of the stack until no more remain and then returns $R$.

\subsection{Maximum Mean Cycle Computation }
\label{sec:cyclescc}

To complete the limit average computation, we need to identify the
maximum mean cycle in a strongly connected component.
We show the adapted algorithm as Alg.~\ref{alg:main-helper-cycle}.

%
%
%
%
%
%
%
%

\begin{algorithm}[tp]
\caption{\small {\bf Computation of the maximum mean weight cycle in
    an SCC.}}
\renewcommand{\utab}{\quad}
\renewcommand{\uln}{\stepcounter{lineno}\>\hspace{-2mm}\thelineno}
\begin{code}
\uln \> {\bf Procedure }  Mean-Cycle-SCC() \\
\uln \> {\bf Input:} $R: S \rightarrow \mathbb{B}(N)$: a symbolic SCC
\\
\uln \> {\bf Output:} $C: \mathbb{B}(N) \rightarrow \mathbb{R}$: a
symbolic maximum \\ 
\hspace*{16em} mean-weight cycle \\
\uln \>\ubegin\\
\uln \> \> Pick $s_o \in S$: an arbitrary initial state \\
\uln \>\> \ufor k =  $0, \ldots, n$ and $v \in S\setminus\{ s_0\}$ \\ 
\uln \> \> \> D[k, $v$, R($v$)] $ \uassign -\infty$ \\
\uln \>\>  D[0, $s_0$, R($s_0$)] $\uassign$ 0 \\
\uln \>\> \ufor k =  $1, \ldots, n$ and $v  \in S$ \\
\uln \> \> \> \ufor $(u, \alpha, v) \in trans$ s.t. $R(u) \neq \bot$
\\
\uln \> \> \> \> $\delta_1 = \gamma((u,v))$ \\
\uln \> \> \> \>\ufor $\delta_2 \in$ domain(D[k, $v$, $\bullet$ ])
\\
\hspace*{11em} and $\delta_3 \in$  domain(D[k-1, $u$, $\bullet$ ]) \\
\uln  \> \>  \> \> \> \uif $\delta_1 \wedge \delta_2  \wedge \delta_3
\not\models \bot$ and \\
\hspace*{8em} D[k-1, $u$, $\delta_3$] $+$ W(($u$,$\alpha$,$v$)) $>$  D[k, $v$, $\delta_2$]   \\
\uln \> \> \> \> \>\>  D[k, $v$, $\delta_2 \wedge \delta_3
\wedge \delta_1$]  $\uassign$ \\
\hspace*{14em} D[k-1, $u$, $\delta_3$] + W(($u$,$\alpha$,$v$)) \\
\uln \> \>  \> \> \>  \>   D[k, $v$, $\delta_2 \wedge \neg(\delta_3 \wedge \delta_1)$] $\uassign$ D[k, $v$, $\delta_2$] \\
\uln \> \>  \> \> \>  \>    Undef D[k, $v$, $\delta_2$]  \\
\uln \> \>  \> \> \> \uendif \\
\uln \>\>  C[R($s_0$)] $\uassign -\infty$ \\
\uln \>\>  \ufor $v  \in S$ \\
\uln \>\>   \> M[$v$, R($v$)] $\uassign +\infty$  \\
\uln \>\>   \> \ufor k = $0, \ldots, n-1$ \\
\uln \>\> \> \> \ufor $\delta_1 \in $ Domain(M[$v$, $\bullet$]),
$\delta_2 \in$ \makebox[0pt][l]{Domain(D[n, $v$, $\bullet$]),} \\
\hspace*{8em} and $\delta_3 \in $  Domain(D[k, $v$, $\bullet$]) \\
\uln \> \>  \> \> \> \uif $\delta_1 \wedge \delta_2  \wedge \delta_3
\not\models \bot$ and \\
\hspace*{8em} M[$v$, $\delta_1$] $>$  (D[n, $v$, $\delta_2$] -  D[k, $v$, $\delta_3$])/(n-k)  \\
\uln \> \> \> \> \> \> M[$v$,$\delta_1\wedge\delta_2\wedge\delta_3$]
$\uassign$ \\ 
\hspace*{12em} (D[n, $v$, $\delta_2$] -  D[k, $v$, $\delta_3$])/(n-k)   \\
\uln  \> \> \> \> \> \>  M[$v$, $\delta_1\wedge \neg(\delta_2 \wedge\delta_3$) ] $\uassign$ M[$v$, $\delta_1$]  \\ 
\uln  \> \> \> \> \> \>  Undef M[$v$, $\delta_1$]   \\
\uln \> \>  \> \> \>  \uendif \\
\uln \>\>  \> \ufor $\delta_1 \in$ Domain(C[$\bullet$]) and $\delta_2 \in$ Domain(M[$v$, $\bullet$]) \\
\uln \>  \> \> \> \uif $\delta_1 \wedge \delta_2  \not\models \bot \; \wedge $  C[$\delta_1$] $<$ M[$v$, $\delta_2$]  \\
\uln \>  \> \> \> \>   C[$\delta_1 \wedge \delta_2$] $\uassign$ M[$v$, $\delta_2$]  \\
\uln \>  \> \> \> \>   C[$\delta_1 \wedge \neg\delta_2$] $\uassign$  C[$\delta_1$]  \\
\uln \>  \> \> \> \>   Undef C[$\delta_1$]  \\
\uln \>  \> \> \> \uendif \\
\uln \>\> \ureturn C
\end{code}
\label{alg:main-helper-cycle}
\end{algorithm}
 
Our algorithm is a feature-aware variant of Karp's original
algorithm~\cite{Karp78}.  As in Karp's algorithm, we chose an
arbitrary initial state $s_0$ and start by computing a function $D$
which for each state $v$ and each $k\in\{ 0,\dotsc, n\}$ gives the
maximal weight of a path of length $k$ from $v$ to $s_0$.  However,
this weight will also depend on the feature guards along paths, so
that $D$ now takes a feature expression as extra input.

After initialization in lines 6-8, computation of $D$ starts in
line~9.  For each pair $k, v$, $D[ k, v]$ is defined on a
\emph{feature partition} of $R( v)$, the feature expression which
governs whether $v$ is present in the current SCC.  Initially (line
7), the domain of $D[ k, v]$ is the coarsest partition of $R( v)$,
which is $R( v)$ itself, and during the iteration in lines 9-17, this
partition is refined as necessary.

For each $k\in\{ 1,\dotsc, n\}$, each $v\in S$, and each transition
$( u, \alpha, v)$, we need to check whether
$D[ k, v]< D[ k- 1, u]+ W(( u, \alpha, v))$, and if it is, update it
to this value.  Now both $D[ k, v]$ and $D[ k- 1, u]$ are defined on
(possibly different) feature partitions, and the transition
$( u, \alpha, v)$ is only enabled for some feature guard $\delta_1$.
Hence we need to find each $\delta_2$ in the domain of $D[ k, v]$ and
each $\delta_3$ in the domain of $D[ k- 1, u]$ for which the
conjunction $\delta_1\land \delta_2\land \delta_3$ is satisfiable
(line 12) and then check whether
$D[ k, v, \delta_2]< D[ k- 1, u, \delta_3]+ W(( u, \alpha, v))$.  If
it is, then $D[ k, v]$ needs to be updated, but only in the part of
its partition where $v$ can be reached from $u$, hence only at
$\delta_1\land \delta_2\land \delta_3$.  That is (lines 14-16), we
need to split the domain of $D[ k, v]$, update the value at
$D[ k, v, \delta_1\land \delta_2\land \delta_3]$, and keep the old
value at $D[ k, v, \delta_2\land \neg( \delta_1\land \delta_3)]$.

In the next part of the algorithm (lines 19-27), we compute
$M[ v]:= \min_{ k< n} \frac{ D[ n, v]- D[ k, v]}{ n- k}$ for each
$v\in S$.  As this again depends on the feature guards on the
transitions, also $M[ v]$ is defined on a feature partition which
initially is set to $R( v)$ (line~20) and then refined as necessary.
Finally, in lines 28-33, we use the same partition refinement
technique once more to compute $C:= \max_{ v\in S} M[ v]$, which per
Karp's theorem~\cite{Karp78} is the maximum mean cycle weight of the
SCC.

\section{Implementation and \\ Evaluation}
\label{se:implex}

We have implemented our algorithms within
ProVeLines, a product line of
verifiers for
SPLs ~\cite{DBLP:conf/splc/CordyCHSL13}.  
ProVeLines
takes as input a specification written in fPromela, a feature-aware extension of the
Promela language~\cite{DBLP:books/daglib/0020982}, which we have
extended further to specify transition weights.
We have modified the code of ProVeLines to include weights on
transitions and to perform family-based and product-based computations
of the maximum mean cycle algorithm.  

As an example, Fig.~\ref{fi:taxi-pml-extract} shows part of our
extended fPromela specification of the taxi-shuttle example from
Section~\ref{se:taxi}. The three features Shuttle, Taxi and License are declared at the beginning of the file. We also show a transition from AirportR (Airport Dropoff location) to AirportP (Airport pick up location ) annotated with a weight of -5. Another available transition in such snippet is from from Pick-Up one location directly to Pick-Up two location; this transition is guarded by feature shuttle and annotated with a weight of 15. 

\begin{figure}[tbp]
  \vspace*{.5ex}
  \centering
  {\ttfamily
    \begin{code}
      typedef features \{ \\
      \> bool Shuttle; \\
      \> bool Taxi; \\
      \> bool License \\
      \}; \\
      features f; \\
      int current = 0; \\
      \dots \\
      active proctype taxi() \{ \\
      \> do :: \\
      \> \> if \> :: (current == AIRPORTR); \\
      \> \> \> \> current = AIRPORTP [-5]; \\
      \> \> \> :: (current == PICKUP1); \\
      \> \> \> \> if \> :: f.Shuttle; \\
      \> \> \> \> \> \> current = PICKUP2 [15]; \\
      \> \> \> \> \> :: !f.Shuttle; \\
      \> \> \> \> \> \> skip; \\
      \> \> \> \> fi; \\
      \> \> \> \dots \\
      \> \> fi; \\
      \> od; \\
      \}
    \end{code}
  }

  \vspace*{-3ex}
  \caption{
    \label{fi:taxi-pml-extract}
    Part of the fPromela specification of the taxi-shuttle example}
\end{figure}

\subsection{Subject Systems}

For testing and experiments, we use a variant of the
taxi-shuttle example from Section~\ref{se:taxi} in which the number of
extra licenses is parameterized.  This variant has $N$ different
extra-license features $L_1,\dotsc, L_N$, each with their own
Pickup-ext$_i$ and Release-ext$_i$ states and transitions that are guarded by the feature
$L_i$.
%
%

As a second case study, we created a  WFTS representing a mine pump
controller, based on an example
in~\cite{Classen:2010:MCL:1806799.1806850}.
The original
example models a system that controlls a water pump to balance
the levels of water and methane in a mine shaft. The  water level needs to
be low for operation; and high levels of methane can lead to an
explosion, and thus  need to be avoided.
The system consists of a water pump, a methane sensor and a command
module that activates or deactivates the water pump.
It is modeled as the parallel composition of five individual FTSs which
model the controller itself, the user (who sends start/stop commands),
the methane alarm, the water sensor, and the methane sensor.

Our WFTS of the mine pump controller has two optional features (the command module
and the methane sensor) and four products; and the annotated the transitions in the main module are annotated with
artificial weights.
The ProveLines encoding of the mine pump controller example has 9441
different states.

\subsection{Results}

\begin{table}[tp]
  \centering
  \caption{
    \label{tbl:taxiresults}
    Maximum limit-average values for the taxi example.}
  \vspace*{.5ex}
  \begin{tabular}{c|c|l}
    product & max. & cycle \\\hline
    $\emptyset$\rule{0pt}{2.5ex} & 12.17 & AP$\to$R2$\to$P2$\to$AR$\to$AP \\
    $\{ L\}$ & 12.17 & AP$\to$R2$\to$P2$\to$AR$\to$AP \\
    $\{ S\}$ & 12.88 & AP$\to$R2$\to$R1$\to$P1$\to$P2$\to$AR$\to$AP \\
    $\{ T\}$ & 14.00 & P1$\to$R2$\to$P2$\to$R1$\to$P1 \\
    $\{ L, S\}$ & 13.30 & AP$\to$R2$\to$R1$\to$Re$\to$ \\
    && \qquad $\to$Pe$\to$P1$\to$P2$\to$AR$\to$AP \\
    $\{ L, T\}$ & 14.00 & P1$\to$R2$\to$P2$\to$R1$\to$P1 \\
    $\{ S, T\}$ & 14.33 & P1$\to$P2$\to$R1$\to$P1 \\
    $\{ L, S, T\}$ & 14.60 & Pe$\to$P1$\to$P2$\to$R1$\to$Re$\to$Pe
  \end{tabular}
\end{table}

\begin{table*}[tp]
  \centering
  \caption{
    \label{tbl:Results}
    Average time consumption of family-based and product-based limit
    average computation on the taxi and the mine pump controller
    examples.}
  \vspace*{.5ex}
  \begin{minipage}{.7\linewidth}
    \renewcommand\thefootnote{\thempfootnote}
    \setlength{\extrarowheight}{1.5pt}
  \begin{tabular}{|l|r|r|rlr|r|}  
    \hline
    \# of features  & \# of products  &  \# of states
    & 
      \multicolumn{3}{c|}{family-based (s)\footnote{mean $\pm$ std.~dev.}} & product-based (s)\footnotemark[\value{mpfootnote}] \\
 \hline\hline
      \multicolumn{7}{c}{Parameterized taxi example} 
    \\ \hline \hline                        
    	3	&	8		& 	52	&	0.25	& 	$\pm$ &	4.57		\% & 0.27		$\pm$ 9.44 \% \\\hline
         4	&	16		&	75	&	0.30	& 	$\pm$ &	3.47		\% & 0.56		$\pm$  1.64 \% \\\hline
         5	&	32		&	98	&	0.44	& 	$\pm$ &	2.99		\% & 1.04		$\pm$  9.04 \% \\\hline
         6	&	64		&	121 &	0.80	& 	$\pm$ &	4.15		\% & 2.19		$\pm$  2.19 \% \\\hline
         7	&	128		& 	144 &	1.83	& 	$\pm$ &	13.24	\% & 4.89		$\pm$  1.36  \% \\\hline
         8	&	256		& 	167 &	3.86	& 	$\pm$ &	1.07		\% & 10.64	$\pm$ 2.01   \%  \\\hline
         9	&	512		& 	190 &	10.84	& $\pm$ &	8.95		\% & 23.25	$\pm$  2.10 \%  \\\hline 
         10	&	1024	& 	213 &	24.63	& $\pm$ &	 6.26		\% & 51.71	$\pm$  1.94  \%  \\\hline
    	11	&	2048	& 	236 &	63.27	& $\pm$ &	5.05		\% &  114.74	$\pm$ 1.79  \%  \\\hline
    	12	&	4096	& 	259 &	142.30	& $\pm$ &	5.27		\% & 251.87	$\pm$ 1.47  \%  \\\hline
    	13	&	8192	& 	282	&	307.56	& $\pm$ &	1.55		\% & 554.16	$\pm$ 1.33  \%  \\\hline	
      \multicolumn{7}{c}{Mine pump controller example} 
    \\ \hline \hline
    	2	&	4	& 9441 	&	291.84	& 	$\pm$ &	2.79		\% & 110.91		$\pm$ 7.61 \% \\\hline	
      \multicolumn{7}{c}{Mine pump controller example,
        optimized\footnote{See Section~\ref{se:discuss}}} 
    \\ \hline \hline
    	2	&	4	& 768 	&		36.01  &	$\pm$  & 7.71		\% & 9.15		$\pm$ 4.33 \% \\\hline	
  \end{tabular}
  \end{minipage}
\end{table*}

Table~\ref{tbl:Results} on the next page shows the running times of
our implementation,
for both family-based and product-based analyses of the parameterized
taxi example and the mine pump controller example. We ran both the
family-based and product-based analyses ten times each.

For the parameterized taxi example, our family-based approach is about
twice as fast as the product-based approach.  For the mine pump
controller however, family-based analysis takes more than twice as
long as product-based analysis.


In table ~\ref{tbl:taxiresults}, we report the complete  results of the maximum limit average for all products of the non-parameterized
taxi example from Section~\ref{se:taxi}.  The third column lists an example cycle in each product with
the given maximum limit average value.  (Here, Pickup-1 is abbreviated as P1 and so on.)

\subsection{Discussion}
\label{se:discuss}

Our results for the parameterized taxi example are as expected.
Symbolic SCCs are shared across products, so that a single computation
over a symbolic SCC can provide answers that can be re-used across
multiple products.  Hence the required time is reduced when using a
family based-approach.


The mine pump controller example has very few products, and SCCs are
generally not shared across products.  Hence the family-based approach
fails to ``lump'' products into families and is, thus, slower than the
product-based analysis.  Additionally, this example has a much larger
state space than the taxi example, hence both product-based and
family-based analysis are quite slow.  Note that the same type of
problems has been reported for this example in other papers, for
example~\cite{Classen:2010:MCL:1806799.1806850}.

By profiling our implementation, we found that computing the maximum
mean cycle for very large symbolic SCCs is taking most of the time in
the family-based approach.
Hence we decided to attempt to perform an abstraction of the mine-pump
controller state space to improve performance.

The mine-pump controller has multiple processes running in parallel.
It is not necessary to consider all possible interleavings of these
processes in order to examine all possible cycles.
Hence we labelled some of its key states as important states and 
considered only transitions between these states.
This abstract model has a much smaller state space, reducing  the running times
for both the family-based and product-based approaches.
Note that the family-based approach is still slower than the
product-based approach, but both are now reasonably fast.


%
%
%
%

We also found that for this mine-pump controller
example, different products induce different sets of finishing times,
and that there is very little sharing of finishing times across products of symbolic SCCs.
Therefore the family-based approach does not improve the performance
for this example, and the overhead introduced by the family-based
approach means it is substantially slower than the product-based
analysis.

\section{Related Work}

\paragraph{Product Line Analysis}

Lauenroth et al.~\cite{Lauenroth:2009:MCD:1747491.1747525}
introduce an algorithm to verify a product line, represented as an I/O
automaton with optional transitions annotated with features, against
properties expressed in computational tree logic (CTL).
Their algorithm checks that every possible I/O automaton that can be
derived satisfies a given CTL property.
Lauenroth et al.\ mention that CTL properties of the form \emph{EG} $f_1$
can be checked by restricting the automaton
and checking if all non-trivial strongly connected components (SCCs)
of this restricted automaton can be reached from the initial state.
They then adapt this algorithm by replacing the computation of SCCs with a procedure to find a path to a cycle,  keeping track of the features required along such a path to a cycle.  
In our case we are instead interested in finding the maximum average cost cycle for each product or set of products. 
Hence we compute  (symbolic) strongly connected components whereas Lauenroth et al. only search for reachable cycles to perform CTL model checking.
Lauenroth et al.\ do not compare the performance of their family-based approach with respect to a product-based approach. 

Classen et al.~\cite{Classen:2013:FTS:2516408.2516486} adapt
the standard algorithm for model checking properties of transition
systems expressed in linear temporal logic (LTL) to analyze a product
line represented as a featured transition system.
Their approach is between 2 and 38 times faster than analyzing each
product individually. Although they represent products symbolically,
they still represent the transition system using explicit states and
transitions.
In subsequent work, Classen \emph{et
  al.}~\cite{DBLP:journals/scp/ClassenCHLS14} extend their approach to
transition systems represented symbolically. They adapt the algorithm
for model checking CTL properties to a family-based approach and show
speed-ups of several orders of magnitude faster than verifying each product
individually.

More recently, Ben-David \emph{et
  al.}~\cite{Ben-David:2015:SMC:2818754.2818780} have adapted
SAT-based model checking of safety properties to a family-based
approach and showed that their approach is substantially faster than
the methods by Classen et al.

\paragraph{Limit-Average Cost}

Quantitative methods are important in performance
analysis~\cite{DBLP:books/daglib/0076234}, reliability
analysis~\cite{DBLP:journals/tc/ShatzWG92}, and other areas of
software engineering.  Long-term average values are often used, for
example to measure mean time between failures or average power
consumption; see
also~\cite{DBLP:journals/ife/Henzinger13,DBLP:journals/computer/HenzingerS07}
for further motivation.

In~\cite{erny:2012:SD:2076807.2077020}, {\v C}ern\'y et al.\
show how limit average cost can be used to measure the distance
between a~specification and an~incorrect implementation.  They define
a limit-average correctness distance to capture how frequently the
specification has to ``cheat'' in order to simulate the incorrect
implementation.  This work is generalized to interfaces and
abstractions
in~\cite{DBLP:conf/popl/CernyHR13,DBLP:journals/tcs/CernyCHR14}.

In~\cite{DBLP:journals/acta/FahrenbergL14,DBLP:journals/tcs/FahrenbergL14},
Fahrenberg and Legay argue more generally for an approach of
quantitative model checking which measures distances between models
and specifications; a similar proposal is Henzinger and
Otop's~\cite{DBLP:conf/concur/HenzingerO13}.  As a specific example,
Boker et al.\ in~\cite{DBLP:journals/tocl/BokerCHK14} extend
LTL with limit-average path accumulation assertions and show that
model checking quantitative Kripke structures with respect to this LTL
extension is decidable.

We are not aware of any family-based analysis methods which compute
the limit average cost for all products in a software product line.



\section{Conclusions and Future Work}

We have introduced a family-based algorithm to compute the maximum
limit average of quality attributes in a software product line.  Our
algorithm is based on symbolic extensions of the standard algorithms
for computing maximum limit average and is able to compute the maximum
limit average of quality attributes for all products in one run.

We have implemented our algorithm by extending ProVeLines, an existing
tool for model checking software product lines, with capabilities to
compute the maximum limit average for product line models annotated
with weights on transitions.
We have used our implementation to evaluate our approach by comparing
the performance of our algorithm against a product-based
(enumerative) approach.

%
%
We have shown that our family-based approach speeds up analysis
compared to a product-based analysis on one class of examples with a
large number of products.  Our approach does not improve performance
for another example system, with more states and fewer products, which
is known to be difficult to analyse using family-based techniques.

There is a number of optimizations which could benefit our
implementation.  Our family-based algorithm can be combined with state
abstraction techniques to scale to large systems.  We could also
consider abstractions over weights to promote a higher level of re-use
when computing the limit-average in a family-based manner.  Finally,
our symbolic representation of strongly connected components can be
made more concise by using a tree-based representation not only of the
set of SCCs but also of each individual SCC.

%
\bibliographystyle{abbrvnat}
\bibliography{doc,bib/rafael,bib/strings-long,bib/proceedings}  
%
%



\end{document}